\documentclass[aps,prd,nofootinbib,twocolumn,superscriptaddress,preprintnumbers]{revtex4}

\usepackage{amssymb}
\usepackage{amsmath}
\usepackage{epsfig}
\usepackage{hyperref}
\usepackage{breakurl}

\usepackage{undertilde}
\makeatletter
\def\simgt{\mathrel{\lower2.5pt\vbox{\lineskip=0pt\baselineskip=0pt
           \hbox{$>$}\hbox{$\sim$}}}}
\def\simlt{\mathrel{\lower2.5pt\vbox{\lineskip=0pt\baselineskip=0pt
           \hbox{$<$}\hbox{$\sim$}}}}
\makeatother

\newcommand{\be}{\begin{equation}}
\newcommand{\ee}{\end{equation}}
\newcommand{\bea}{\begin{eqnarray}}
\newcommand{\eea}{\end{eqnarray}}
\newcommand{\Eq}[1]{Eq.~(\ref{#1})}

\newcommand{\Fig}[1]{Fig.~\ref{#1}}

\newcommand{\mPl}{m_{\rm Pl}}

\newcommand{\mNLSP}{m_{\rm NLSP}}

\newcommand{\NLSP}{{\rm NLSP}}

\begin{document}


\title{Gravitino Freeze-In}

\author{Clifford Cheung}
\affiliation{Berkeley Center for Theoretical Physics, 
  University of California, Berkeley, CA 94720, USA}
\affiliation{Theoretical Physics Group, 
  Lawrence Berkeley National Laboratory, Berkeley, CA 94720, USA}

\author{Gilly Elor}
\affiliation{Berkeley Center for Theoretical Physics, 
  University of California, Berkeley, CA 94720, USA}
\affiliation{Theoretical Physics Group, 
  Lawrence Berkeley National Laboratory, Berkeley, CA 94720, USA}

\author{Lawrence Hall}
\affiliation{Berkeley Center for Theoretical Physics, 
  University of California, Berkeley, CA 94720, USA}
\affiliation{Theoretical Physics Group, 
  Lawrence Berkeley National Laboratory, Berkeley, CA 94720, USA}

\begin{abstract}
We explore an alternative mechanism for the production of gravitino dark matter whereby relic gravitinos originate from the decays of superpartners which are still in thermal equilibrium, i.e.~via freeze-in.  Contributions to the gravitino abundance from freeze-in can easily dominate over those from thermal scattering over a broad range of parameter space, e.g.~when the scalar superpartners are heavy.  Because the relic abundance from freeze-in is independent of the reheating temperature after inflation, collider measurements may be used to unambiguously reconstruct the freeze-in origin of gravitinos.  In particular, if gravitino freeze-in indeed accounts for the present day dark matter abundance, then the lifetime of the next-to-lightest superpartner is uniquely fixed by the superpartner spectrum.
\end{abstract}

\maketitle

\section{Introduction}

Supersymmetry is an elegant and well-motivated extension of the standard model which solves the hierarchy problem and carries extensive  phenomenological consequences.  Despite its successes, however, supersymmetry suffers from an assortment of cosmological difficulties which are referred to collectively as the ``cosmological gravitino problem'', which has two components:
\begin{itemize}
\item[a)] Late decaying superpartners can produce electromagnetic or hadronic radiation that can adversely affect Big Bang Nucleosynthesis (BBN).
\item[b)] Relic gravitinos produced by scattering and decaying superpartners can overclose the universe.
\end{itemize}
The gravitino problem highlights a tension between supersymmetry and cosmology which is highly robust.  This is so because the existence of the gravitino is required by supergravity, and because the couplings of the gravitino to superpartners are uniquely fixed by soft masses and a single additional parameter, the gravitino mass.

At the same time, there are a number of approaches by which to address these issues.  For example, a) may be evaded if the scale of supersymmetry breaking is very high, as in anomaly mediation, in which case the gravitino will decay safely before BBN.  Alternatively, superpartner decays to the gravitino can be made sufficiently rapid if the scale of supersymmetry breaking is low or intermediate, as in gauge mediation.

\begin{figure}[t]
\begin{center} 
\includegraphics[scale=0.8]{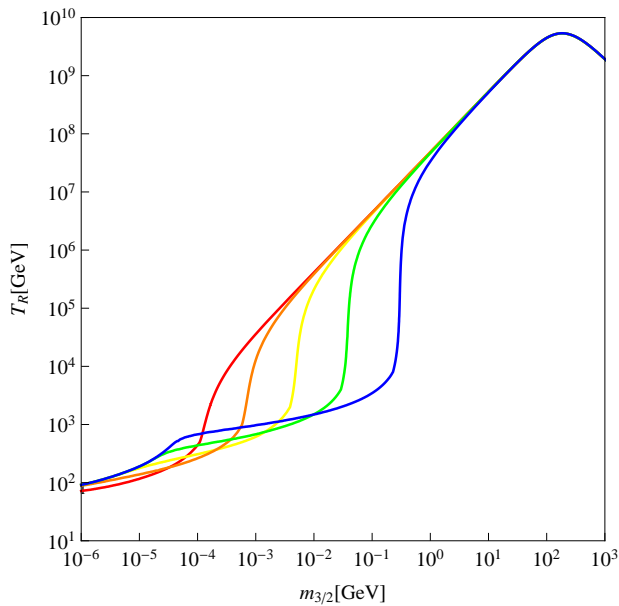}
\end{center}
\caption{Contours of $\Omega_{3/2} h^2 = 0.11$ for gaugino masses fixed to
$\{ m_{\tilde b}, m_{\tilde w}, m_{\tilde g}\}
= \{100, 210, 638\} \textrm{ GeV}$.  The \{red, orange, yellow, green, blue\} contours correspond to 
universal scalar masses \{500 GeV, 1 TeV, 2 TeV, 4 TeV, 8 TeV\}.}
\label{fig:changesquark}
\end{figure}

Likewise, b) can be resolved if $m_{3/2} \lesssim $ keV, in which case the gravitino is simply too light to overclose the universe\footnote{However, note that this class of theories is in tension with warm dark matter constraints.}.  For $m_{3/2} \gtrsim $ keV, the authors of \cite{Moroi:1993mb} famously showed that b) can be avoided if the reheating temperature after inflation, $T_{R}$, is below a critical value which depends on the superpartner spectrum and is shown in Fig.~\ref{fig:changesquark}.   Numerous new physics proposals---for example, ones including new stable charged particles or superweakly interacting particles---suffer from an analogous overclosure problem which may be evaded by appropriately lowering $T_R$.

The conventional wisdom is that low $T_R$ is disfavored; for example, $T_R \gtrsim 10^9$ GeV for high scale leptogenesis \cite{leptoTR}. As a consequence, the vast majority of papers on gravitino cosmology have focused on the portion of Fig.~\ref{fig:changesquark} at high $T_R$.   However, the cosmological baryon asymmetry can be generated at much lower temperatures, for example via soft leptogenesis \cite{Grossman:2004dz}, and in this case gravitinos become a virtue rather than a problem: not only are both a) and b) resolved, but gravitinos can fully account for the observed dark matter for a wide range of masses,
as depicted in Fig.~\ref{fig:changesquark}.

In this paper, we investigate a more or less overlooked regime of gravitino  cosmology corresponding to the vertical incline in critical $T_R$ shown as a function of $m_{3/2}$ in \Fig{fig:changesquark}.  Here gravitino dark matter arises dominantly from the freeze-in mechanism, which was studied in some generality in \cite{FIMP}.  In this setup, a feebly interacting dark matter particle is produced via the decays of particles which are still in thermal equilibrium.  Crucially, since the decay rates of these particles fix the final abundance of dark matter, the associated lifetimes are hence constrained by the observed dark matter abundance.  Furthermore, because the production is dominated at low temperatures, the freeze-in abundance is largely independent of $T_R$, explaining the vertical incline in critical $T_R$.  For the case of gravitino freeze-in, decays of superpartners in thermal equilibrium produce a final yield of
\bea
Y_{3/2}^{\rm decay} &\simeq& \frac{405}{2\pi^4} \sqrt{\frac{5}{2}} \frac{\mPl}{g_*^{3/2}} \sum_i \frac{\Gamma_i }{ m_i^2},
\label{eq:FI}
\eea
where $i$ sums over all superpartners, $m_i$ and $\Gamma_i$ are superpartner masses and partial decay widths to the gravitino, and $\mPl$ is the reduced Planck mass\footnote{Throughout, sums over superpartners will implicitly include a degeneracy factor---for instance, a factor of 8 for gluinos, etc.}.

As we will see, the freeze-in abundance of gravitino dark matter depends solely on the superpartner spectrum and $m_{3/2}$, a quantity which is straightforwardly inferred from the mass and lifetime of the NLSP when it decays to the gravitino LSP.  
Thus, for a given superpartner spectrum, the constraint of $\Omega_{3/2}^{\rm decay} h^2 \simeq 0.11$ entirely fixes the lifetime of the NLSP.
Because these quantities are experimentally accessible, we chance upon the rather amazing prospect of reconstructing the origin of gravitino dark matter through collider measurements.  
 For example, for degenerate heavy squarks and gluinos at a mass $m$, the NLSP lifetime is
\bea
\tau_{\NLSP}  \simeq 10^{-7}  \textrm{ sec} \, \left( \frac{\mNLSP}{300 \textrm{ GeV}} \right) \left( \frac{m}{\mNLSP} \right)^6,
\label{eq:tauNLSP}
\eea
if gravitino freeze-in accounts for the present day abundance of dark matter. 
Note that this proposal is a specific instance of the generalized cosmological scenario discussed in \cite{Cheung:2010gj,Cheung:2010gk}.

\begin{figure}[t]
\begin{center} 
\includegraphics[scale=0.8]{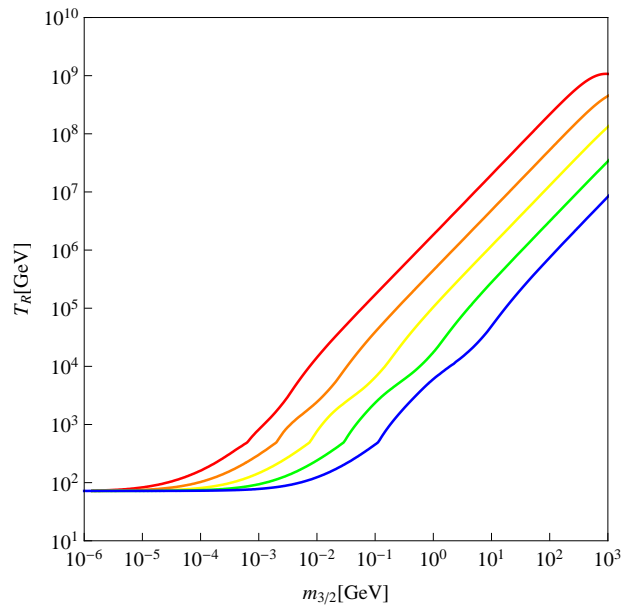}
\end{center}
\caption{Contours of $\Omega_{3/2} h^2 = 0.11$ for universal
scalar masses fixed to 500 GeV.  The \{red, orange, yellow, green, blue\} contours correspond to a bino mass $m_{\tilde b}=$ \{500 GeV, 1 TeV, 2 TeV, 4 TeV, 8 TeV\}, where $m_{\tilde w}$ and $m_{\tilde g}$ are fixed assuming gaugino mass unification at $M_{\rm GUT}\sim 10^{16}$ GeV.}
\label{fig:changegaugino}
\end{figure}

While the lifetime $\tau_\NLSP$ indicated by \Eq{eq:tauNLSP} is effectively long-lived on collider time scales, a number of theoretical and experimental collaborations have suggested that the LHC is capable of measuring the long-lived decays of the sizable number of charged or colored metastable NLSPs which will typically slow and eventually stop within the detector material.  Sufficiently long lifetimes can easily arise in theories of split supersymmetry  \cite{Arvanitaki:2005nq}, as well as theories with very weakly coupled particles like gravitinos  \cite{Buchmuller:2004rq}, axinos  \cite{Brandenburg:2005he}, goldstini  \cite{Cheung:2010mc,Cheung:2010qf}, sterile sneutrinos \cite{seesaw}, and dark matter \cite{Feng:2004mt,Cheung:2010gk}. 
Hence, stopped NLSPs allow for a range of $10^{-9}-10^6$ sec to be probed in early LHC running, and indeed bounds on stopped gluinos have already been set by the CMS collaboration \cite{Khachatryan:2010uf}.  At higher luminosities, neutral NLSPs might also be probed if their lifetimes lie in the range $10^{-9} - 10^{-5}$ sec.  As such, gravitino freeze-in offers a novel mechanism of dark matter generation which has direct implications for the LHC in the near term.
  
\section{Gravitino Cosmology}

Assuming that the messenger scale of supersymmetry breaking is below the Planck scale, then
the gravitino is the lightest of all the superpartners and is thus an attractive R-parity stabilized dark matter candidate.   Typically, the gravitino mass is considered in the range keV $\lesssim m_{3/2} \lesssim 1$ GeV, where the lower bound arises from warm dark matter constraints and the upper bound arises from tension with BBN\footnote{The quantitative BBN bound on $m_{3/2}$ varies with the nature and mass of the NLSP.  Moreover, in some cases it can be evaded altogether, e.g.~with sneutrino NLSP or R-parity violation.}.  
Broadly speaking, gravitinos are produced via three distinct physical mechanisms, each with a much different dependence on the reheating temperature after inflation, $T_R$, and the gravitino mass, $m_{3/2}$.   

\begin{figure}[t]
\begin{center} 
\includegraphics[scale=0.8]{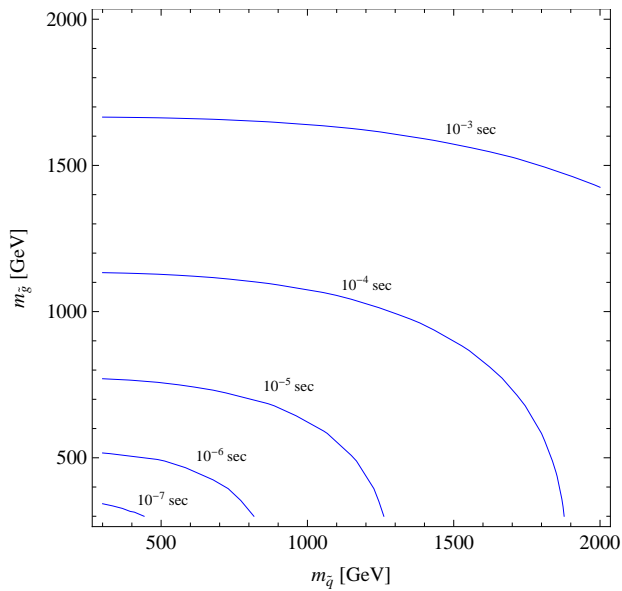}
\end{center}
\caption{Prediction of $\tau_{\NLSP}$ for gravitino dark matter arising from freeze-in, for $\mNLSP$ = 300 GeV.  The sum over superpartner contributions is assumed to be dominated by gluinos and degenerate squarks as shown in Eq. (\ref{eq:NLSPlifetime}).}
\label{fig:tau}
\end{figure}

\subsection{Modes of Production}
First, there is a contribution to the gravitino abundance arising from NLSPs which freeze-out and then decay to the gravitino, as in the so-called superWIMP scenario \cite{superWIMP}.  This contribution is highly model dependent and can easily be negligible
since the final relic gravitino abundance is down by a factor of $m_{3/2} / m_{\rm NLSP}$ relative to the freeze-out abundance of the NLSP, which can itself be small if the NLSP has strong annihilation channels.
In addition, BBN is in tension with the superWIMP mechanism as the origin of the dark matter \cite{Kawasaki:2008qe}.   For these reasons, we will ignore superWIMP contributions and focus on other gravitino production mechanisms.

A second source of gravitino production is the thermal scattering of superpartners in the early universe.  Since the goldstino couples to gauginos through a dimension five operator, the associated scattering processes are dominated at high temperatures, and so the final abundance of gravitinos depends linearly on $T_R$.  For example, the yield of gravitinos from gaugino scattering goes parametrically as
\bea
Y_{3/2}^{\textrm{scatt}} &\propto& \frac{T_R \sum_a  g_a^2  m_a^2}{m_{3/2}^2 \mPl},
\eea
where $a=1,2,3$ sums over the gauge group and $m_a$ are the gaugino masses.
 This scattering contribution has been the primary focus of existing work on gravitino dark matter, and corresponds to the 
straight, sloped portions of the contours in Figs.~\ref{fig:changesquark} and \ref{fig:changegaugino}, which depict contours of the total gravitino abundance $\Omega_{3/2} h^2 = 0.11$ in the $(m_{3/2}, T_R)$ plane for different choices for the superpartner spectra.

Lastly,  gravitinos may be produced by freeze-in: that is, from the decays of superpartners which are still in thermal equilibrium.  The near vertical portions of the curves in Figs.~\ref{fig:changesquark} and \ref{fig:changegaugino} correspond to freeze-in, which is infrared dominated and thus independent of $T_R$.  Plugging in for $\Gamma_i$ in \Eq{eq:FI} yields the parametric dependence
\bea
Y_{3/2}^{\textrm{decay}} &\propto& \frac{\sum_i m_i^3}{m_{3/2}^2 \mPl},
\eea
where $i$ sums over all superpartners. 
As $T_R$ drops below the superpartner masses, the superpartners are not efficiently produced from reheating, and hence gravitino production arises from the  exponentially tiny Boltzmann tail fixing the abundance of superpartners. 

The size of the freeze-in region in Figs.~\ref{fig:changesquark} and \ref{fig:changegaugino} is determined by the competition between $Y_{3/2}^{\rm scatt}$ and $Y_{3/2}^{\rm decay}$.  Thus, let us define $T_R^*$ to be the reheating temperature at which these two quantities are equal.  Clearly, the range in $T_R$ in which freeze-in is operative runs from the superpartner masses up to $T_R^*$. Since this cross-over value of the reheating temperature goes as 
\bea
T_R^* &\propto& \frac{\sum_i m_i^3}{\sum_a m_a^2},
\label{eq:crossover}
\eea
this implies that the freeze-in region will diminish for larger gaugino masses, and will grow for larger scalar masses.  

The trend implied by \Eq{eq:crossover} is verified in \Fig{fig:changesquark}, which depicts $\Omega_{3/2} h^2 = 0.11$ from a total gravitino yield of $Y_{3/2} = Y_{3/2}^{\rm scatt} +Y_{3/2}^{\rm decay}$ in the $(m_{3/2}, T_R)$ plane.  One can see that  the freeze-in region grows with increasing scalar masses.  This is the case because heavier scalars imply larger decay rates without commensurately larger contributions from thermal scattering at high temperatures.
For the allowed region of gravitino masses, $ 1 \textrm{ keV} \lesssim m_{3/2} \lesssim  1\textrm{ GeV}$, the reheating temperature required for gravitino dark matter varies over $100 \textrm{ GeV}  \lesssim T_R \lesssim 10^7\textrm{ GeV}$.  

We now stress a key point: {\it a sizable fraction of this range corresponds to gravitino freeze-in and is thus insensitive to $T_R$}, which is fortuitous because $T_R$ is not an experimentally accessible quantity.  Hence this allows for the unique possibility of reconstructing the freeze-in origin of gravitino dark matter from LHC measurements.   As shown in \Fig{fig:changesquark}, for squarks accessible at LHC ($m_{\tilde q} \lesssim 2 \textrm{ TeV}$) this occurs in about $30\% - 50\%$ of the logarithmic  range of $T_R$, while for heavier squarks the range is even greater.  From a theoretical standpoint, it is straightforward to make the scalars quite heavy while keeping the gauginos light with an R-symmetry.
Note that while \Fig{fig:changesquark} was produced assuming degenerate scalar masses, an almost identical plot results if the top and bottom squarks are pushed down to near the weak scale, as considered in \cite{split squarks}.  

\Fig{fig:changegaugino} also verifies the trend indicated by \Eq{eq:crossover}, since it shows the freeze-in region diminishing for increasing gaugino masses.  This is the case because heavier gauginos imply larger scattering cross-sections at high energies and thus a larger contribution arising from $Y_{3/2}^{\rm scatt}$.  From a top-down viewpoint, theories with very heavy gauginos and light scalars are difficult to accommodate, since very large values of $m_{\tilde g}$ tend to drag up $m_{\tilde q}$ and exacerbate fine-tuning of electroweak symmetry breaking.    Thus, we conclude that it is actually theoretically difficult to obtain theories in which the freeze-in region is small, and so a large freeze-in region is a typical feature of many reasonable models.

Also, let us not that immediately after freeze-in, the produced gravitinos are highly relativistic, but they become non-relativistic as the temperature drops below their mass, yielding cold dark matter.   At the same time, very light gravitinos, $m_{3/2} < 10$ keV, yield warm dark matter, especially since freeze-in arises from decays of superpartners in the exponential tail of their thermal distribution.

\subsection{Reconstructing the Origins of Dark Matter}

Let us now consider the extent to which the freeze-in origin of gravitino dark matter might actually be verified at the LHC. Assuming that the present day abundance of gravitinos arises entirely from freeze-in, $Y_{3/2} = Y_{3/2}^{\rm decay}$, one can rewrite  \Eq{eq:FI} as
\bea
 m_{3/2} Y_{3/2} &=& \frac{0.26}{g_*^{3/2} }\sqrt{ \frac{ \mNLSP}{\tau_\NLSP}} \, \sum_i  \left( \frac{m_i}{\mNLSP} \right)^3,
 \label{eq:FI2}
\eea
which is obtained from 
\bea
\tau_\NLSP^{-1} &=& \frac{1}{48 \pi} \frac{m_\NLSP^5}{m_{3/2}^2 \mPl^2}
\eea
while normalizing the partial widths of the superpartners decaying into gravitinos, $\Gamma_i$, with respect to the NLSP decay width, $\Gamma_\NLSP$, so $\Gamma_i /m_i^5= \Gamma_\NLSP /m_\NLSP^5$.
From \Eq{eq:FI2} we see that $m_{3/2} Y_{3/2} \propto \sum_i m_i^3$ is dominated by the very heaviest superpartners.  Assuming that the superpartner spectrum is measured, \Eq{eq:FI2} can be inverted to yield a critical prediction for the NLSP lifetime in terms of the superpartner spectrum:
\bea
\tau_{\NLSP} &=& 4\times 10^{17} \textrm{ GeV}^{-2} \times   \frac{m_\NLSP}{g_*^3}\left[\sum_i\left( \frac{m_i}{m_\NLSP}\right)^3 \right]^2 \nonumber\\
&\simeq & 7 \times 10^{-5}  \textrm{ sec} \times \left(\frac{150}{g_*} \right)^3 \, \left( \frac{300 \, \mbox{GeV}}{\mNLSP} \right)^5 \nonumber  \\
 && \left[ \frac{9}{11}\left( \frac{m_{\tilde{q}}}{\textrm{TeV}} \right)^3 + \frac{2}{11}\left( \frac{m_{\tilde{g}}}{\textrm{TeV}} \right)^3 \right]^2.
\label{eq:NLSPlifetime}
\eea
The second line corresponds to an approximation in which the gravitinos are produced dominantly by squarks and gluinos.
  Also, while the value of $g_*$ actually varies substantially with temperature, for this approximate expression we have normalized its value to 150, which lies somewhere between the $g_*$ for the standard model and the minimal supersymmetric standard model.
The corresponding prediction obtained by numerically solving the Boltzmann equations is shown in \Fig{fig:tau}.  

Thus, $\tau_\NLSP$ may be as short as $10^{-7}$ sec for a squashed supersymmetric spectrum and as long as 100 sec, the approximate bound from BBN, for an extremely split spectrum.  Note that this entire range of lifetimes is relatively long-lived on the time scales relevant to collider physics.

Fortunately, some fraction of metastable charged or colored NLSPs, such as the squark, slepton, chargino, or gluino, will interact with and eventually stop within the material of the LHC detectors.  A number of groups evaluated this stopping efficiency, as well as prospects for performing precision spectroscopic measurements on the ensuing late NLSP decays \cite{colliderNLSP}.  Recently a search for stopped gluinos performed by the CMS collaboration placed a bound of $m_{\tilde g} < 398 \textrm{ GeV}$ for a stable gluino \cite{Khachatryan:2010uf}.  More generally, it is expected that CMS will effectively probe lifetimes of stopped particles in the range of $10^{-6}-10^6$ sec \cite{CMS-PAS EXO-09-001}.  

In the case of neutral NLSPs, such as the neutralino or sneutrino, stopping will not occur.  That said, a sizeable fraction $L / \gamma c\tau_\NLSP$ of events may still decay within the length of the detector, $L \simeq {\cal O}(1 \textrm{ m})$, allowing for the possibility of lifetime probes in the range $10^{-9} - 10^{-5}$ sec at high luminosity.

The key point is that there exists a precision correlation between the superpartner spectrum and the NLSP lifetime which, if verified, would provide very strong evidence for gravitino dark matter arising from freeze-in.   Also, while $T_R$ cannot be inferred accurately from collider measurements, precisely because freeze-in occurs on the near vertical part of the contours in \Fig{fig:changesquark} and \Fig{fig:changegaugino}, the reheating temperature will have a very strong upper bound at the order of magnitude level.

\section{Conclusions}

If the messengers of supersymmetry breaking are below the Planck scale, then the gravitino is the LSP and is thus a prime candidate for dark matter.  We find that for a large range in $T_R$, gravitino dark matter is predominantly produced by freeze-in and is thus insensitive to $T_R$.  Furthermore, there is a direct correlation between the cosmological abundance of dark matter and the decay rate of the NLSP to gravitinos.  The NLSP lifetime, given in \Eq{eq:NLSPlifetime} and shown as contours in \Fig{fig:tau}, allows for a precision test at the LHC of the freeze-in origins of gravitino dark matter.  Moreover, the reheat temperature can then be inferred, at least to within a couple of orders of magnitude, from \Fig{fig:changesquark}.

  In \cite{axinoprob} we will provide additional motivations for gravitino freeze-in through an investigation of the cosmology of the QCD axino, the supersymmetric partner of the QCD axion. Our discussion will center on a ``QCD axino problem'' which is entirely analogous to the gravitino problem but which occurs in a complementary region of $m_{3/2}$.  Together, the combined axino and gravitino problem completely exclude the possibility of a high $T_R$ and thus much of parameter space in which gravitino production arises from thermal scattering.  This indicates a robust tension between the axion solution to the strong CP problem and supersymmetry which strongly favors a low reheating temperature.

\begin{center}
{\bf Acknowledgments}
\end{center}
We would like to thank Piyush Kumar for collaboration during the early stages of this work. This work was supported in part by the Director, Office of Science,  Office of High Energy and Nuclear Physics, of the US Department of Energy under Contract DE-AC02-05CH11231and by the National Science Foundation on grant PHY-0457315.

\end{document}